\documentclass{llncs}

\usepackage[latin1]{inputenc}
\usepackage[T1]{fontenc}
\usepackage{amsmath,latexsym,amssymb}
\usepackage{epsf}
\usepackage{fancybox}
\usepackage{graphicx}
\usepackage{url}
\newdimen\proofrulebreadth \proofrulebreadth=.05em
\newdimen\proofdotseparation \proofdotseparation=1.25ex
\newdimen\proofrulebaseline \proofrulebaseline=2ex
\newcount\proofdotnumber \proofdotnumber=3
\let\then\relax
\def\hfi{\hskip0pt plus.0001fil}
\mathchardef\squigto="3A3B
\newif\ifinsideprooftree\insideprooftreefalse
\newif\ifonleftofproofrule\onleftofproofrulefalse
\newif\ifproofdots\proofdotsfalse
\newif\ifdoubleproof\doubleprooffalse
\let\wereinproofbit\relax
\newdimen\shortenproofleft
\newdimen\shortenproofright
\newdimen\proofbelowshift
\newbox\proofabove
\newbox\proofbelow
\newbox\proofrulename
\def\shiftproofbelow{\let\next\relax\afterassignment\setshiftproofbelow\dimen0 }
\def\shiftproofbelowneg{\def\next{\multiply\dimen0 by-1 }%
\afterassignment\setshiftproofbelow\dimen0 }
\def\setshiftproofbelow{\next\proofbelowshift=\dimen0 }
\def\setproofrulebreadth{\proofrulebreadth}

\def\prooftree{
\ifnum  \lastpenalty=1
\then   \unpenalty
\else   \onleftofproofrulefalse
\fi
\ifonleftofproofrule
\else   \ifinsideprooftree
        \then   \hskip.5em plus1fil
        \fi
\fi
\bgroup
\setbox\proofbelow=\hbox{}\setbox\proofrulename=\hbox{}%
\let\justifies\proofover\let\leastoX\proofoverdots\let\Justifies\proofoverdbl
\let\using\proofusing\let\[\prooftree
\ifinsideprooftree\let\]\endprooftree\fi
\proofdotsfalse\doubleprooffalse
\let\thickness\setproofrulebreadth
\let\shiftright\shiftproofbelow \let\shift\shiftproofbelow
\let\shiftleft\shiftproofbelowneg
\let\ifwasinsideprooftree\ifinsideprooftree
\insideprooftreetrue
\setbox\proofabove=\hbox\bgroup$\displaystyle 
\let\wereinproofbit\prooftree
\shortenproofleft=0pt \shortenproofright=0pt \proofbelowshift=0pt
\onleftofproofruletrue\penalty1
}

\def\eproofbit{
\ifx    \wereinproofbit\prooftree
\then   \ifcase \lastpenalty
        \then   \shortenproofright=0pt  
        \or     \unpenalty\hfil         
        \or     \unpenalty\unskip       
        \else   \shortenproofright=0pt  
        \fi
\fi
\global\dimen0=\shortenproofleft
\global\dimen1=\shortenproofright
\global\dimen2=\proofrulebreadth
\global\dimen3=\proofbelowshift
\global\dimen4=\proofdotseparation
\global\count255=\proofdotnumber
$\egroup  
\shortenproofleft=\dimen0
\shortenproofright=\dimen1
\proofrulebreadth=\dimen2
\proofbelowshift=\dimen3
\proofdotseparation=\dimen4
\proofdotnumber=\count255
}

\def\proofover{
\eproofbit 
\setbox\proofbelow=\hbox\bgroup 
\let\wereinproofbit\proofover
$\displaystyle
}%
\def\proofoverdbl{
\eproofbit 
\doubleprooftrue
\setbox\proofbelow=\hbox\bgroup 
\let\wereinproofbit\proofoverdbl
$\displaystyle
}%
\def\proofoverdots{
\eproofbit 
\proofdotstrue
\setbox\proofbelow=\hbox\bgroup 
\let\wereinproofbit\proofoverdots
$\displaystyle
}%
\def\proofusing{
\eproofbit 
\setbox\proofrulename=\hbox\bgroup 
\let\wereinproofbit\proofusing
\kern0.3em$
}

\def\endprooftree{
\eproofbit 
  \dimen5 =0pt
\dimen0=\wd\proofabove \advance\dimen0-\shortenproofleft
\advance\dimen0-\shortenproofright
\dimen1=.5\dimen0 \advance\dimen1-.5\wd\proofbelow
\dimen4=\dimen1
\advance\dimen1\proofbelowshift \advance\dimen4-\proofbelowshift
\ifdim  \dimen1<0pt
\then   \advance\shortenproofleft\dimen1
        \advance\dimen0-\dimen1
        \dimen1=0pt
        \ifdim  \shortenproofleft<0pt
        \then   \setbox\proofabove=\hbox{%
                        \kern-\shortenproofleft\unhbox\proofabove}%
                \shortenproofleft=0pt
        \fi
\fi
\ifdim  \dimen4<0pt
\then   \advance\shortenproofright\dimen4
        \advance\dimen0-\dimen4
        \dimen4=0pt
\fi
\ifdim  \shortenproofright<\wd\proofrulename
\then   \shortenproofright=\wd\proofrulename
\fi
\dimen2=\shortenproofleft \advance\dimen2 by\dimen1
\dimen3=\shortenproofright\advance\dimen3 by\dimen4
\ifproofdots
\then
        \dimen6=\shortenproofleft \advance\dimen6 .5\dimen0
        \setbox1=\vbox to\proofdotseparation{\vss\hbox{$\cdot$}\vss}%
        \setbox0=\hbox{%
                \advance\dimen6-.5\wd1
                \kern\dimen6
                $\vcenter to\proofdotnumber\proofdotseparation
                        {\leaders\box1\vfill}$%
                \unhbox\proofrulename}%
\else   \dimen6=\fontdimen22\the\textfont2 
        \dimen7=\dimen6
        \advance\dimen6by.5\proofrulebreadth
        \advance\dimen7by-.5\proofrulebreadth
        \setbox0=\hbox{%
                \kern\shortenproofleft
                \ifdoubleproof
                \then   \hbox to\dimen0{%
                        $\mathsurround0pt\mathord=\mkern-6mu%
                        \cleaders\hbox{$\mkern-2mu=\mkern-2mu$}\hfill
                        \mkern-6mu\mathord=$}%
                \else   \vrule height\dimen6 depth-\dimen7 width\dimen0
                \fi
                \unhbox\proofrulename}%
        \ht0=\dimen6 \dp0=-\dimen7
\fi
\let\doll\relax
\ifwasinsideprooftree
\then   \let\VBOX\vbox
\else   \ifmmode\else$\let\doll=$\fi
        \let\VBOX\vcenter
\fi
\VBOX   {\baselineskip\proofrulebaseline \lineskip.2ex
        \expandafter\lineskiplimit\ifproofdots0ex\else-0.6ex\fi
        \hbox   spread\dimen5   {\hfi\unhbox\proofabove\hfi}%
        \hbox{\box0}%
        \hbox   {\kern\dimen2 \box\proofbelow}}\doll%
\global\dimen2=\dimen2
\global\dimen3=\dimen3
\egroup 
\ifonleftofproofrule
\then   \shortenproofleft=\dimen2
\fi
\shortenproofright=\dimen3
\onleftofproofrulefalse
\ifinsideprooftree
\then   \hskip.5em plus 1fil \penalty2
\fi
}

\newcommand{\tensor}[0]{\langle *\rangle}
\newcommand{\comp}[0]{>\!\!\!>\!\!\!>}
\newcommand{\eee}[0]{$\&\!\!\!\&\!\!\!\&$}
\newcommand{\sem}[1]{[\!\![#1]\!\!]}
\newcommand{\bind}[0]{>\!\!\!>\!\!\!=}

\begin{document}

\title{The Arrow Calculus as a Quantum Programming Language}

\author{Juliana Kaizer Vizzotto\inst{1}, Andr\'e Rauber Du Bois\inst{2} 
and Amr Sabry\inst{3}}

\institute{Mestrado em Nanoci\^encias, Centro Universit\'ario Franciscano\\
 Santa Maria, RS/ Brazil
\and
PPGI, Universidade Cat\'olica de Pelotas \\
Pelotas, RS/Brazil
\and  Department of Computer Science, Indiana University\\
 Bloomington, USA}

\maketitle

\begin{abstract}
We express quantum computations (with measurements) using the arrow calculus
extended with monadic constructions. This framework expresses quantum
programming using well-understood and familiar classical patterns for
programming in the presence of computational effects.  In addition, the five
laws of the arrow calculus provide a convenient framework for
\emph{equational reasoning} about quantum computations that include
measurements.
%
%
\end{abstract}

\section{Introduction}

\emph{Quantum} computation~\cite{NielsenChuang00} can be understood as a
\emph{transformation} of information encoded in the state of a \emph{quantum}
physical system.  Its basic idea is to encode data using quantum bits
(qubits).  Differently from the classical bit, a qubit can be in a
\emph{superposition} of basic states leading to ``quantum parallelism.'' This
form of parallelism is due to the non-local wave character of quantum
information and is qualitatively different from the classical notion of
parallelism. This characteristic of quantum computation can greatly increase
the processing speed of algorithms.  However, quantum data types are
computationally very powerful not only due to superposition. There are other
odd properties like \emph{measurement}, in which the observed part of the
quantum state and every other part that is \emph{entangled} with it
immediately lose their wave character.

These interesting properties have led to the development of very efficient
quantum algorithms, like Shor's quantum algorithm for factorizing
integers~\cite{Shor94}, and Grover's quantum search on
databases~\cite{grover96fast}.  Another important theme is the development of
quantum cryptographic techniques~\cite{bennett93teleporting}.

Since these discoveries, much research has been done on quantum
computation. Summarizing the field of research we can classify it according
three main areas: i) physical implementations of quantum computers, ii)
development of new quantum algorithms; and iii) design of quantum programming
languages.

This work is about the design of a quantum programming language, and
consequently about a high-level, structured and well-defined way to develop
new quantum algorithms and to \emph{reason} about them.

We have been working on semantic models for quantum programming.  In previous
work~\cite{viz06} we established that general quantum computations (including
measurements) are an instance of the category-theoretic concept of
arrows~\cite{hughes:arrows}, a generalization of \emph{monads}~\cite{moggi89}
and \emph{idioms}~\cite{mcbride08}. Translating this insight to a practical
programming paradigm has been difficult however. On one hand, directly using
arrows is highly non-intuitive, requiring programming in the so-called
``point-free'' style where intermediate computations are manipulated without
giving them names. Furthermore reasoning about arrow programs uses nine,
somewhat idiosyncratic laws. 


In recent work, Lindley \emph{et. al.}~\cite{lindley:tac} present the
\emph{arrow calculus}, which is a more friendly version of the original
presentation of arrows.  The arrow calculus augment the simply typed lambda
calculus with four constructs satisfying five laws. Two of these constructs
resemble function abstraction and application, and satisfy familiar beta and
eta laws.  The remaining two constructs resemble the unit and bind of a
monad, and satisfy left unit, right unit, and associativity laws.  Basically,
using the arrow calculus we can understand arrows through classic well-known
patterns.

In this work we propose to express quantum computations using the arrow
calculus axtended with monadic constructions. We show that quantum 
programming can be expressed using well-understood and familiar classical 
patterns for programming in the presence of computational effects. 
Interestingly, the five laws of the arrow
calculus provide a convenient framework for \emph{equational reasoning} about
quantum computations (including measurements). 

This work is organized as follows. The next two sections review the
background material on modeling quantum computation using classical arrows.
Section~\ref{sec:arrowc} presents the \emph{arrow calculus}. We show the
quantum arrow calculus in Section~\ref{sec:qarrowc}. We express some
traditional examples of quantum computations using the quantum calculus.
Additionally, we illustrate how we can use the calculus to reason about
quantum programs.  Section~\ref{sec:conclusion} concludes with a discussion
of some related works.  Finally, Appendix~\ref{app:lambda} presents the
constructs of simply-typed lambda calculus, Appendix~\ref{app:monadic} gives
an extension of the simply-typed lambda calculus with monadic constructions,
and Appendix~\ref{app:qcomp} reviews general quantum computations.

\vspace{-0.4cm}
\section{Classic Arrows}
\label{sec:arrows}

The simply-typed lambda calculus is an appropriate model of pure functional
programming (see Appendix~\ref{app:lambda}). The standard way to model
programming in the presence of effects is to use
\emph{monads}~\cite{moggi91notions} (see Appendix~\ref{app:monadic}). Arrows, like monads, are used to
elegantly program notions of computations in a pure functional setting. But
unlike the situation with monads, which wrap the \emph{results of
computations}, arrows wrap the \emph{computations} themselves.

From a programming point of view, \emph{classic arrows} extend the
simply-typed lambda calculus with one type and three constants satisfying
nine laws (see Figure~1). The type $A \leadsto B$ denotes a computation that
accepts a value of type $A$ and returns a value of type $B$, possibly
performing some side effects. The three constants are: $\mathit{arr}$, which
promotes a function to a pure arrow with no side effects; \comp, which
composes two arrows; and $\mathit{first}$, which extends an arrow to act on
the first component of a pair leaving the second component unchanged.

To understand the nine equations, we use some auxiliary functions. The function
$\mathit{second}$, is like $\mathit{first}$, but acts on the second component
of a pair, and $f \eee g$, applies arrow $f$ and $g$ to the same argument and
then pairs the results.
\vspace{-0.5cm}
\begin{figure}[!htb]
\label{fig:classica}
\caption{Classic Arrows}
\[
\begin{array}{l}\hline
\textbf{Types}\\
\multicolumn{1}{l}{
\begin{array}{l}
\;\;arr :: (A \to B) \to (A \leadsto B)\\
\;\;(\comp) :: (A \leadsto B) \to (B \leadsto C) \to (A \leadsto C) \\
\;\;first :: (A \leadsto B) \to (A \times C \leadsto B \times C)
\end{array}}
\\ 
\textbf{Definitions} \\
\begin{array}{l}
\;\;second : (A \leadsto B) \to (C \times A \leadsto C \times B) \\
\;\;second = \lambda f . arr \; swap \comp first \; f  \comp arr \; swap \\
\;\;(\eee) : (C \leadsto A) \to (C \leadsto B) \to (C \leadsto A \times B)\\ 
\;\;(\eee) = \lambda f . \lambda g . arr \; sup \comp first \; f \comp second \; g
\end{array}
\\\textbf{Equations}\\
\begin{array}{llcl}
\;\;(\leadsto_1) & arr \; id \comp f &=& f\\
\;\;(\leadsto_2) & f \comp arr \; id&=& f\\
\;\;(\leadsto_3)& (f \comp g) \comp h &=& f \comp (g \comp h) \\
\;\;(\leadsto_4)& arr (g . f) &=& arr \; f \comp arr \; g\\
\;\;(\leadsto_5)& first (arr \; f)&=& arr (f \times id)\\
\;\;(\leadsto_6)& first (f \comp g) &=& first \; f \comp first \; g\\
\;\;(\leadsto_7)& first \;f \comp arr(id \times g) &=&  arr (id \times g) \comp first \; f \\
\;\;(\leadsto_8)& first \; f \comp arr \; fst &=& arr \; fst \comp f\\
\;\;(\leadsto_9)& first (first \; f) \comp arr &=& arr \; assoc \comp first \; f\\
\end{array}
\\\hline
\end{array}
\]
\end{figure}

\vspace{-0.4cm}
\section{Quantum Arrows}
\label{sec:qarrows}

Quantum computation is generally expressed in the framework of a Hilbert
space (see Appendix~\ref{app:qcomp} for a short review of that model). As
expressive and as convenient is this framework for mathematical reasoning, it
is not easily amenable to familiar programming techniques and
abstractions. In recent work~\cite{viz06} however, we established that this
general model of quantum computations (including measurements) can be
structured using the category-theoretic concept of arrows.  Figure~2 explains
the main ideas which we elaborate on in the remainder of this section.

In the figure, we have added type definitions (i.e, type synonyms) for
convenience. Type $\mathbf{Vec} \; A$ means that a vector is a function
mapping elements from a vector space orthonormal basis to complex numbers
(i.e., to their probability amplitudes).  Type $\mathbf{Lin}$ represents a
linear operator (e.g, a unitary matrix) mapping a vector of type $A$ to a
vector of type $B$. Note that if we \emph{uncurry} the arguments $A$ and $B$,
it turns exactly into a square matrix (i.e, $\mathbf{Vec} \; (A,B))$.  Type
$\mathbf{Dens} \;A$ stands for density matrices and it is straight to build from
$\mathbf{Vec}$.  Type $\mathbf{Super} \; A \; B$ means a superoperator mapping a
density matrix of type A to a density matrix of type B. This type can be
understood by interpreting it in the same style as $\mathbf{Lin}$.

\vspace{-0.5cm}
\begin{figure}[!htb]
\label{fig:quantuma}
\caption{Quantum  Arrows}
\[
\begin{array}{l}\hline
\textbf{Type Definitions}\\ 
\multicolumn{1}{l}{
\begin{array}{lcl}
\;\;type \;\mathbf{Vec} \; A  &=& A \to \mathbb{C}\\
\;\;type \; \mathbf{Lin} \;A \;B &=& A \to \mathbf{Vec} \; B\\ 
\;\;type \; \mathbf{Dens} \;A &=& \mathbf{Vec}\; (A,A)\\
\;\;type \; \mathbf{Super} \; A \; B& =& (A,A) \to \mathbf{Dens} \; B\\
\end{array}}\\
\textbf{Syntax}\\ \begin{array}{llcl}
\;\;\text{Types} & A,B,C& ::=& ... \;\mathbf{Vec} \; A \; | \; 
\mathbf{Lin} \; A \; | \; \mathbf{Dens}\; A \;|\; \mathbf{Super} \; A \;B \\
\;\;\text{Terms}& L, M, N & ::= & ...\; |\; return \;
|\; \bind \;|\; arr\; | \; \comp \;|\; first\\
\end{array}\\ 
\textbf{Monadic Definitions} \\
\begin{array}{l}
\;\;return : A \to \mathbf{Vec} \;A\\
\;\;return \;a \;b = \mathsf{if}\; a == b \;\mathsf{then} \; 1.0 \; \mathsf{else} \; 0.0\\  
\;\;(\bind): \mathbf{Vec} \; A \to (A \to \mathbf{Vec}\; B) \to \mathbf{Vec}
 \; B\\
\;\;va \bind f = \lambda b. \sum a \;(va \; a) (f \; a \; b)\\
\end{array}
\\ 
\textbf{Auxiliary Definitions} \\
\begin{array}{l}
\;\;fun2lin : (A \to B) \to \mathbf{Lin} \; A \; B  \\
\;\;fun2lin \;f  = \lambda \; a .return \; (f \; a)\\
\;\;(\tensor) : \mathbf{Vec} \; A \to \mathbf{Vec} \; B \to \mathbf{Vec} \;(A,B)\\
\;\;v_1 \tensor v_2 = \lambda \;(a,b). v_1 \; a \;*\; v_2 \; b\\
\end{array}
\\ 
\textbf{Arrow Types and Definitions} \\
\begin{array}{l}
\;\;arr : (A \to B) \to \mathbf{Super} \; A \; B \\
\;\;arr \;f = fun2lin \; (\lambda \; (b_1, b_2) \to (f\; b_1, f\; b_2))\\ 
\;\;(\comp) :: (\mathbf{Super} \;A \; B) \to (\mathbf{Super} \; B \; C) \to (\mathbf{Super} A \; C) \\
\;\;f \comp g = \lambda \; b . (f \; b \bind g)\\
\;\;first :: (\mathbf{Super} \;A \; B) \to (\mathbf{Super} \; (A \times C) \; (B \times C))\\
\;\;first \; f \; ((b_1,d_1),(b_2,d_2)) = permute \;((f (b_1,b_2)) \tensor \;return \; (d_1,d_2))\\
\;\;\;\;\;\;\; \mathsf{where} \;permute \; v \;((b_1,b_2),(d_1,d_2)) = v \;((b_1,d_1),(b_2,d_2))\\
\end{array}
\\\hline
\end{array}
\]
\end{figure}
\vspace{-0.5cm}

We have defined in our previous work~\cite{viz06} the arrow operations for
quantum computations into two levels. First we have proved that \emph{pure}
quantum states (i.e, vector states) are an instance of the concept of
monads~\cite{moggi89}. The definitions of the monadic functions are shown in
Figure~2. The function $\mathit{return}$ specifies how to
construct vectors and $\bind$ defines the behavior of an application of
matrix to a vector.  Moreover we have used the auxiliary functions
$\mathit{fun2lin}$, which converts a classical (reversible) function to a
linear operator, and $\tensor$ which is the usual tensor product in vector
spaces.

The function $\mathit{arr}$ constructs a quantum superoperator from a pure
function by applying the function to both vector and its dual. The
composition of arrows just composes two superoperators using the monadic
\emph{bind}. The function $\mathit{first}$ applies the superoperator $f$ to
the first component (and its dual) and leaves the second component unchanged.

We have proved in our previous work that this superoperator instance of
arrows satisfy the required nine equations~\cite{viz06}.

\section{The Arrow Calculus}
\label{sec:arrowc}

In this section we present the arrow calculus~\cite{lindley:tac} and show the
translation of the calculus to classic arrows (described in
Section~\ref{sec:arrows}) and vice versa.  The translation is important
because it essentially corresponds to the denotational semantic function for
the quantum version of the arrow calculus. The material of this section 
closely follows the original presentation in~\cite{lindley:tac}.

\subsection{The Calculus}
\label{subsec:acalculus}

The arrow calculus as shown in Figure~3 extends the core lambda
calculus with four constructs satisfying five laws.
\vspace{-0.5cm}
\begin{figure}[!htb]
\label{acalculus}
\caption{Arrow Calculus}
\[
\begin{array}{l}\hline
\textbf{Syntax}\\ \begin{array}{llcl}
\;\;\text{Types} & A,B,C& ::=& \ldots\; |\; A \leadsto B\\
\;\;\text{Terms}& L, M, N & ::= & \ldots \;|\; \lambda^{\bullet}x. Q\\
\;\;\text{Commands}& P,Q, R& ::=& L \bullet P \; | \;[M] \;| \;\mathsf{let}\; 
x = P \; \mathsf{in} \; Q\\
\end{array}\\ 
\textbf{Types}\\
\begin{array}{cc}\\
\prooftree
\Gamma;x :A \vdash Q ! B
\justifies
\Gamma \vdash \lambda^{\bullet}x.Q:A\leadsto B
\endprooftree&
\;\;\;\;\;\;\prooftree
\Gamma \vdash L : A \leadsto B \;\;\;\; \Gamma; \Delta \vdash M : A
\justifies
\Gamma; \Delta \vdash L \bullet M ! B
\endprooftree
\end{array} \\ \\
\begin{array}{cc}
\;\;\;\;\;\;\;\;\prooftree
\Gamma, \Delta \vdash M :A
\justifies
\Gamma; \Delta \vdash [M]!A
\endprooftree&
\;\;\;\;\;\;\;\;\;\prooftree
\Gamma; \Delta \vdash P ! A \;\;\;\; \Gamma; \Delta, x:A \vdash Q! B
\justifies
\Gamma; \Delta \vdash \mathsf{let} \; x = P\; \mathsf{in} \; Q ! B  
\endprooftree
\end{array}
\\ \textbf{Laws}\\
\begin{array}{llcl}
\;\;(\beta^{\leadsto}) & (\lambda^{\bullet} x. Q)\bullet M &=& Q[x:=M]\\
\;\;(\eta^{\leadsto}) & \lambda^{\bullet} x. (L \bullet [x]) &=& L\\
\;\;(\text{left})&  \mathsf{let} \; x = [M] \; \mathsf{in} \; Q&=& Q[x:=M]\\
\;\;(\text{right})& \mathsf{let} \; x = P \; \mathsf{in} \; [x] &=& P\\
\;\;(\text{assoc})& \mathsf{let} \; y = (\mathsf{let} 
\; x = P \; \mathsf{in} \; Q) \; \mathsf{in} \; R &=& \mathsf{let} 
\; x = P \; \mathsf{in} \; (\mathsf{let} 
\; y = Q \; \mathsf{in} \; R)\\
\end{array}
\\\hline
\end{array}
\]
\end{figure}
\vspace{-0.5cm}
Type $A \leadsto B$ denotes a computation that accepts a value of type 
$A$ and returns a value of type $B$, possibly performing some side 
effects.

There are two syntactic categories. Terms are ranged over by $L,M,N$, and
commands are ranged over by $P,Q,R$. In addition to the terms of the core
lambda calculus, there is one new term form: arrow abstraction
$\lambda^{\bullet}x.Q$.  There are three command forms: arrow application $L
\bullet M$, arrow unit $[M]$ (which resembles unit in a monad), and arrow
bind $\mathsf{let} \; x = P \; \mathsf{in} \;Q$ (which resembles bind in a
monad).
 
In addition to the term typing judgment $\Gamma \vdash M : A$ there is also a
command typing judgment $\Gamma ; \Delta \vdash P ! A$.  An important feature
of the arrow calculus is that the command type judgment has two environments,
$\Gamma$ and $\Delta$, where variables in $\Gamma$ come from ordinary lambda
abstractions $\lambda x. N$, while variables in $\Delta$ come from arrow
abstraction $\lambda^{\bullet}x.Q$.

Arrow abstraction converts a command into a term.  Arrow abstraction closely
resembles function abstraction, save that the body $Q$ is a command (rather
than a term) and the bound variable $x$ goes into the second environment
(separated from the first by a semicolon).

Conversely, arrow application, $ L \bullet M ! B$ embeds a term into a
command.  Arrow application closely resembles function application. The arrow
to be applied is denoted by a term, not a command; this is because there is
no way to apply an arrow that is itself yielded. This is why there are two
different environments, $\Gamma$ and $\Delta$: variables in $\Gamma$ may
denote arrows that are applied to arguments, but variables in $\Delta$ may
not.

Arrow unit, $[M] ! A$, promotes a term to a command.  Note that in the
hypothesis there is a term judgment with one environment (i.e, there is a
comma between $\Gamma$ and $\Delta$), while in the conclusion there is a
command judgment with two environments (i.e, there is a semicolon between
$\Gamma$ and $\Delta$).

Lastly, using \texttt{let}, the value returned by a command may be bound.

Arrow abstraction and application satisfy beta and eta laws,
$(\beta^{\leadsto})$ and $(\eta^{\leadsto})$, while arrow unit and bind
satisfy left unit, right unit, and associativity laws, (left), (right), and
(assoc). The beta law equates the application of an abstraction to a bind;
substitution is not part of beta, but instead appears in the left unit
law. The (assoc) law has the usual side condition, that $x$ is not free in
$R$.

\vspace{-0.3cm}
\subsection{Translation}

The translation from the arrow calculus to classic arrows, shown
below, gives a denotational semantics for the arrow calculus.
\vspace{-0.1cm}
\[
\begin{array}{lcl}
\sem{ \lambda^{\bullet}x.Q} &=& \sem{Q}_x\\
\sem{L \bullet M}_{\Delta} &=& arr (\lambda \Delta.\sem{M}) \comp \sem{L}\\
\sem{[M]}_{\Delta} &=& arr (\lambda \Delta. \sem{M})\\
\sem{ \mathsf{let} \; x = P\; \mathsf{in} \; Q}_{\Delta} &=& (arr\; id \;\eee \;\sem{P}_{\Delta}) 
\comp \sem{Q}_{\Delta,x}\\
\end{array}
\]
\vspace{-0.1cm}
\noindent An arrow calculus term judgment $\Gamma \vdash M : A$ maps into a classic
arrow judgment $\Gamma \vdash \sem{M} : A$, while an arrow calculus command
judgment $\Gamma; \Delta \vdash P!A$ maps into a classic arrow judgment
$\Gamma \vdash \sem{P}_{\Delta} : \Delta \leadsto A$. Hence, the denotation
of a command is an arrow, with arguments corresponding to the environment
$\Delta$ and result of type $A$.

We omitted the translation of the constructs of core lambda calculus as they
are straightforward homomorphisms. The translation of the arrow abstraction
$\lambda^{\bullet}x.Q$ just undoes the abstraction and call the
interpretation of $Q$ using $x$. Application $L \bullet P$ translates to
\comp, $[M]$ translates to $\mathit{arr}$ and $\mathsf{let} \; x = P\;
\mathsf{in} \; Q$ translates to pairing \eee (to extend the environment with
$P$) and composition \comp (to then apply $Q$).

The inverse translation, from classic arrows to the arrow calculus is defined
as:
\vspace{-0.2cm}
\[
\begin{array}{lcl}
\sem{arr}^{-1} &=& \lambda f.\lambda^{\bullet} x.[f\;x]\\
\sem{(\comp)}^{-1} &=& \lambda f.\lambda g. \lambda^{\bullet} x. g \bullet (f \bullet x)\\
\sem{first}^{-1} &=& \lambda f. \lambda^{\bullet} z . \mathsf{let} \; x = f \bullet \mathsf{fst} \;z \; 
\mathsf{in} \; [(x,\mathsf{snd}\;z)]\\
\end{array}
\]
\noindent Again we omitted the translation of the constructs of core lambda
calculus as they are straightforward homomorphisms. Each of the three
constants from classic arrows translates to an appropriate term in the arrow
calculus.

\section{The Arrow Calculus as a Quantum Programming Language}
\label{sec:qarrowc}

In this section we discuss how the arrow calculus can be used as a quantum
programming language. 

We start by showing quantum programs using the standard quantum circuit
notation. The lines carry quantum bits.  The values flow from left to right
in steps corresponding to the alignment of the boxes which represent quantum
gates.  Gates connected via bullets to another wire are called
\emph{controlled operations}, that is, the wire with the bullet conditionally
controls the application of the gate.  The circuit in
Figure~4 represents a quantum program for the Toffoli
gate.
\vspace{-0.5cm}
\begin{figure}[h]
\label{fig:tof-circuit}
\caption{Circuit for the Toffoli gate}
\begin{center}
\scalebox{0.6}{\includegraphics{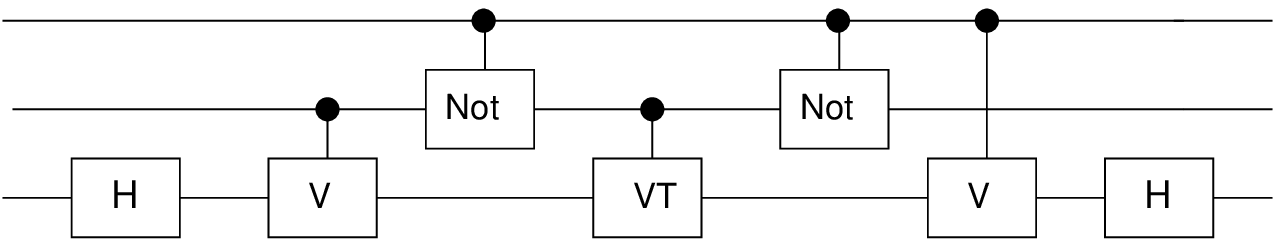}}
\end{center}
\end{figure}
\vspace{-0.4cm}
Using the classic arrows approach for quantum programming presented in
Section~\ref{sec:qarrows} and using the type of booleans, $\mathbf{Bool}$, as the
orthonormal basis for the qubit, this program would be codded as follows:
\[
\begin{array}{ll}
\mathsf{toffoli} ::& \mathbf{Super} \;(\mathbf{Bool},\mathbf{Bool},\mathbf{Bool}) 
\;(\mathbf{Bool},\mathbf{Bool},\mathbf{Bool}) \\
\mathsf{toffoli} = &arr \;(\lambda (a_0, b_0, c_0) \to (c_0, (a_0, b_0))) \comp\\
          &(first \;H  \comp arr \;(\lambda (c_1, (a_0, b_0)) \to ((b_0, c_1), a_0))) \comp\\
          &(first \;cV \comp arr \;(\lambda ((b_1, c_2), a_0) \to ((a_0, b_1), c_2))) \comp\\
          &(first \;cNot \comp arr \;(\lambda ((a_1, b_2), c_2) \to ((b_2, c_2), a_1))) \comp ...
\end{array}
\]
\noindent
As already noted by Paterson~\cite{PatersonRA:notation} this notation is
cumbersome for programming.  This is a ``point-free'' notation, rather
different from the usual way of writing functional programs, with $\lambda$
and $\mathsf{let}$. Paterson introduced syntactic sugar for arrows, which we
have used in our previous work~\cite{viz06}.  However, the notation simply
abbreviates terms built from the three constants, and there is no claim about
reasoning with arrows.  Using the \emph{quantum arrow calculus} presented in
Figure~\ref{qacalculus}, this program would be like:

\[
\begin{array}{ll}
\mathsf{toffoli} ::& \mathbf{Super} \;(\mathbf{Bool},\mathbf{Bool},\mathbf{Bool}) 
\;(\mathbf{Bool},\mathbf{Bool},\mathbf{Bool}) \\
\mathsf{toffoli} = &  \lambda^{\bullet} . (x,y,z). \mathsf{let} \; z' = H \bullet z \; \mathsf{in} \\
&\;\; \mathsf{let} \; (y', z'') = cV \bullet (y,z') \; \mathsf{in}\\
&\;\;\; \mathsf{let} \; (x',y'') = cNot  \bullet (x,y') \mathsf{in} \ldots \\
\end{array}
\]


This style is more convenient and elegant as it is very similar to the usual
familiar classical functional programming and is amenable to formal reasoning
in a convenient way.  Consider, for instance, the program which applies the
quantum \textsf{not} gate twice.  That is obviously equivalent to
identity. To do such a simple proof using the \emph{classic arrows} we need
to learn how to use the \emph{nine} arrow laws and also to recover the
definitions of the functions $\mathit{arr}$, $\comp$ and $\mathit{first}$ for
quantum computations presented in Figure~2.

The action of the quantum not gate, \textsf{QNot}, is to swap the amplitude
probabilities of the qubit. For instance, \textsf{QNot} applied to
$|0\rangle$ returns $|1\rangle$, and vice versa.  But \textsf{QNot} applied
to $\alpha |0\rangle +\beta |1\rangle $ returns $\alpha |1\rangle +\beta
|0\rangle $.

Given the classical definition of \textsf{not} as follows:
\[
\mathsf{not} = \lambda x. \mathsf{if} \; x == True \; \mathsf{then} \; False \; 
\mathsf{else} \; True: \;\mathbf{Bool} \to \mathbf{Bool}
\]
\noindent Using the arrow calculus, the \textsf{QNot} would be written as:
\[
\mathsf{QNot} =  \lambda^{\bullet} y.[\mathsf{not} \;y] : \; \mathbf{Super} \; \mathbf{Bool} \; \mathbf{Bool}.
\]
\noindent Then, the program which applies the \textsf{QNot} twice, would be:
\[
\Gamma \vdash  \lambda^{\bullet} x.\mathsf{let}\; w = (\lambda^{\bullet} z.[\mathsf{not} \;z]) 
\bullet x \; \mathsf{in} \;(\lambda^{\bullet} y.[\mathsf{not} \;y]) \bullet w
\]
\noindent
Again the syntax, with arrow abstraction and application, resembles lambda
calculus.  Now we can use the intuitive arrow calculus laws (from Figure~3)
to prove the obvious equivalence of this program with identity. The proof
follows the same style of the proofs in classical functional programming.
\[
\begin{array}{ll}
\lambda^{\bullet} x.\mathsf{let}\; w = (\lambda^{\bullet} z.[\mathsf{not} \;z]) 
\bullet x \; \mathsf{in} \;(\lambda^{\bullet} y.[\mathsf{not} \;y]) \bullet w \; &=^{(\beta^{\leadsto})}\\
\lambda^{\bullet} x.\mathsf{let}\; w = [\mathsf{not} \;x] \; 
\mathsf{in} \;(\lambda^{\bullet} y.[\mathsf{not} \;y]) \bullet w  &=^{(\text{left})}\\
\lambda^{\bullet} x.  (\lambda^{\bullet} y.[\mathsf{not} \;y]) \bullet 
(\mathsf{not} \;x)  &=^{(\beta^{\leadsto})}\\
\lambda^{\bullet} x.  [\mathsf{not} (\mathsf{not}\; x)]  &=^{def. not}\\
\lambda^{\bullet} x.  [x]&
\end{array}
\]

It is interesting to note that we have two ways for defining
superoperators. The first way is going directly from classical functions to
superoperators as we did above for $not$, using the default definition of
$\mathit{arr}$.  The other way is going from the monadic pure quantum
functions to superoperators.  As monads are a special case of
arrows~\cite{hughes:arrows} there is \emph{always} a translation from monadic
functions to arrows.  Hence, any $\mathbf{Lin} \;A \;B$ is a special case of
$\mathbf{Super} \; A\; B$.

\begin{figure}[!htb]
\label{qacalculus}
\caption{Quantum Arrow Calculus}
\[
\begin{array}{l}\hline
\textbf{Syntax}\\ \begin{array}{llcl}
\text{Types} & A,B,C& ::=& \ldots\;| \; \mathbf{Bool}\;|\;\mathbf{Dens} \; A\;|\; 
\mathbf{Vec} \;A \; |\; \mathbf{Super} \; A \;B\\
\text{Terms}& L, M, N & ::= &  \; 
[ T ] \; | \; \mathsf{let} \; 
x = M \; \mathsf{in} \; N|\; \lambda^{\bullet}x. Q \; | \;\mathsf{+}\;
|\; \mathsf{-}\\
\text{Commands}& P,Q, R& ::=& L \bullet P \; | \;[M] \;
| \;\mathsf{let}\; 
x = P \; \mathsf{in} \; Q \; | \; \mathsf{meas}\; |\; \mathsf{trL}\\
\end{array}\\ 
\textbf{Monad Types}\\
\begin{array}{cc}\\
\prooftree
\Gamma \vdash M : A
\justifies
\Gamma \vdash [M]: \mathbf{Vec}\; A
\endprooftree&
\;\;\;\;\prooftree
\Gamma \vdash M : \mathbf{Vec} \; A\;\;\;\;\Gamma, x : A \vdash N : \mathbf{Vec} \; B
\justifies
\Gamma \vdash \mathsf{let} \; x = M \; \mathsf{in} \; N : \mathbf{Vec}\; B
\endprooftree
\end{array}\\
\begin{array}{cc}\\
\prooftree
\Gamma \vdash M,N : \mathbf{Vec}\; A
\justifies
\Gamma \vdash M \mathsf{+} N: \mathbf{Vec}\; A
\endprooftree \;\;&
\prooftree
\Gamma \vdash M,N : \mathbf{Vec}\; A
\justifies
\Gamma \vdash M \mathsf{-} N: \mathbf{Vec}\; A
\endprooftree
\end{array}  \\
\textbf{Arrow Types}\\
\begin{array}{cc}\\
\prooftree
\Gamma;x :A \vdash Q ! \;\mathbf{Dens} \; B
\justifies
\Gamma \vdash \lambda^{\bullet}x.Q:\mathbf{Super} \; A\; B
\endprooftree&
\;\;\;\;\;\;\prooftree
\Gamma \vdash L : \mathbf{Super} \;A \; B \;\;\;\; \Gamma; \Delta \vdash M :\;A
\justifies
\Gamma; \Delta \vdash L \bullet M ! \; \mathbf{Dens} \; B
\endprooftree
\end{array} \\ \\
\begin{array}{cc}
\;\;\;\;\;\;\;\;\prooftree
\Gamma, \Delta \vdash M :A
\justifies
\Gamma; \Delta \vdash [M]! \;\mathbf{Dens} \;A
\endprooftree&
\;\;\;\;\;\;\;\;\;\prooftree
\Gamma; \Delta \vdash P ! \; \mathbf{Dens} \;A \;\;\;\; \Gamma; \Delta, x:A \vdash Q!\; \mathbf{Dens}\; B
\justifies
\Gamma; \Delta \vdash \mathsf{let} \; x = P\; \mathsf{in} \; Q ! \; \mathbf{Dens}\;B  
\endprooftree
\end{array}\\
\begin{array}{cc}\\
\;\;\;\;\;\;\;\;\prooftree
\justifies
\Gamma;x:A \vdash \mathsf{meas} \;!\;\mathbf{Dens}\;(A,A)
\endprooftree&
\;\;\;\;\;\;\;\;\;\prooftree
\justifies
\Gamma;x:(A,B) \vdash \mathsf{trL} \;!\; \mathbf{Dens} \; B
\endprooftree
\end{array}
\\\\\hline
\end{array}
\]
\end{figure}

Hence, we construct the quantum arrow calculus in Figure~5 in three levels.
First we inherit all the constructions from simply-typed lambda calculus 
with the type of booleans and with classical \textsf{let} and  \textsf{if}
(see Appendix~\ref{app:lambda}).  Then we
add the monadic \emph{unit}, $[\;]$, to build pure vectors (over booleans),
$\mathsf{let}$ to sequence computations with vectors, and plus and minus to
add and subtract vectors (the monadic calculus~\cite{moggi89} with its laws
is presented in Appendix~\ref{app:monadic}).  Finally, we add the
constructions of the arrow calculus. The appeal of using the arrows approach
is because we can express measurement operations (i.e, extract classical
information from the quantum system) inside the formalism. Therefore, we have
two computations for measurements on mixed states, $\mathsf{meas}$ and
$\mathsf{trL}$. The computation $\mathsf{meas}$ returns a classical value and 
a post-measurement state of the quantum
system. The computation $\mathsf{trL}$ \emph{traces out} or
\emph{projects} part of the quantum state (the denotation of these
operations is provided in Appendix~\ref{app:measop}).

To exemplify the use of the monadic constructions, consider, for example, the
\textsf{hadamard} quantum gate, which is the source of superpositions.  For
instance, \textsf{hadamard} applied to $|0\rangle$ returns $|0\rangle +
|1\rangle$, and applied to $|1\rangle$ returns $|0\rangle - |1\rangle $. But,
\textsf{hadamard} applied to $|0\rangle + |1\rangle$ returns $|0\rangle$, as
it is a reversible gate. To define this program in the quantum arrow
calculus, we just need to define its work for the basic values, $|0\rangle$
and $|1\rangle$, as follows:
\[
\begin{array}{ll}
\mathsf{hadamard} = \lambda x. \mathsf{if} \; x == True \; & \mathsf{then} \; 
[False] - [True] \\
&\mathsf{else} \; [False] + [True] : \mathbf{Lin} \;\mathbf{Bool} \; \mathbf{Bool}\\
\end{array}
\]
\noindent
Then, the superoperator  would be:
\[
\mathsf{Had} =  \lambda^{\bullet} y.[ \mathsf{hadamard}\;y] : \mathbf{Super} \;\mathbf{Bool} \;\mathbf{Bool}
\]

Another interesting class of operations are the so-called \emph{quantum
controlled operations}. For instance, the controlled not, \textsf{Cnot},
receives two qubits and applies a not operation on the second qubit depending
on the value of the first qubit. Again, we just need to define it for the
basic quantum values:
\[
\begin{array}{ll}
\mathsf{cnot} =  \lambda (x,y). \mathsf{if} \; x \;& \mathsf{then}\; [(x, \mathsf{not} \;y)] \;\\ 
&\mathsf{else}\; [(x,y)]:\mathbf{Lin}\; (\mathbf{Bool}, \mathbf{Bool}) \; (\mathbf{Bool}, \mathbf{Bool})
\end{array}
\]
\noindent
Again, the superoperator of type $ \mathbf{Super} 
\;(\mathbf{Bool}, \mathbf{Bool}) \;(\mathbf{Bool},\mathbf{Bool})$ would be
$\mathsf{Cnot} =  \lambda^{\bullet} (x,y).[ \mathsf{cnot} \;(x,y)]$.

The motivation of using superoperators is that we can express \emph{measurement}
operations inside of the formalism. One classical example of quantum
algorithm which requires a measurement operation is the quantum
teleportation~\cite{bennett93teleporting}.  It allows the transmission of a
qubit to a partner with whom is shared an entangled pair. Below we define 
the two partners of a teleportation algorithm.

\[
\begin{array}{l}
\mathsf{Alice} : \mathbf{Super} \; (\mathbf{Bool},\mathbf{Bool}) \; (\mathbf{Bool},\mathbf{Bool})\\
\begin{array}{ll}
\mathsf{Alice} = \lambda^{\bullet} (x,y). & \mathsf{let} \; (x',y') = \mathsf{Cnot} \bullet (x,y) \; 
\mathsf{in} \;\\
&\;\; \mathsf{let}\; q =   ( \mathsf{Had} \bullet x', y') \; \mathsf{in} \;\\
&\;\;\;\;\mathsf{let} \; (q',v) = \mathsf{meas} \bullet q \; \mathsf{in} \; 
\mathsf{trL} \bullet (q,v)
\end{array}
\end{array}
\]

\[
\begin{array}{l}
\mathsf{Bob} : \mathbf{Super} \; (\mathbf{Bool},\mathbf{Bool},\mathbf{Bool}) \; \mathbf{Bool}\\
\begin{array}{ll}
\mathsf{Bob} = \lambda^{\bullet} (x,y,z). & \mathsf{let} \; (z',x') = \mathsf{Cnot} \bullet (z,x) \; 
\mathsf{in} \;\\
&\;\; \mathsf{let} \; (y',x'') =   ( \mathsf{Cz} \bullet (y,x')) \; \mathsf{in} \; 
\mathsf{trL} \bullet ((y',z'),x'')
\end{array}
\end{array}
\]

\section{Conclusion}
\label{sec:conclusion}

We have presented a lambda calculus for general quantum programming that
builds on well-understood and familiar programming patterns and reasoning
techniques. Besides supporting an elegant functional programming style for
quantum computations, the quantum arrow calculus allows reasoning about
\emph{general} or \emph{mixed} quantum computations.  This is the first work
proposing reasoning about \emph{mixed} quantum computations. 
The equations of the arrow calculus plus the equations of the monadic calculus 
provide indeed a powerful mechanism to make proofs about quantum programs.  
In~\cite{alt07}
we have proposed very similar reasoning techniques, however for \emph{pure}
quantum programs.  Also, in~\cite{tonder04} the author presents a quantum lambda
calculus based on linear logic, but just for pure quantum computations.

\section*{Acknowledgements}
We thank Jeremy Yallop for very helpful comments.

\bibliographystyle{splncs}
\bibliography{teste}

\appendix

\section{Simply-Typed Lambda Calculus}
\label{app:lambda}

The simply-typed lambda calculus with the type of booleans, 
and with \textsf{let} and  \textsf{if} is shown in Figure~6.
Let $A, B, C$ range over types, $L,M,N$ range over
terms, and $\Gamma, \Delta$ range over environments. A type judgment $\Gamma
\vdash M : A$ indicates that in environment $\Gamma$ term $M$ has type
$A$. As presented in the arrow calculus~\cite{lindley:tac}, we are using a
Curry formulation, eliding types from terms.

\begin{figure}[!htb]
\label{fig:lambda}
\caption{Simply-typed Lambda Calculus}
\[
\begin{array}{l}\hline
\textbf{Syntax}\\\begin{array}{llcl}
\text{Types} & A,B,C& ::=& \mathbf{Bool}\;|\; A \times B\; |\; A \to B\\
\text{Terms}& L, M, N & ::= & x \;|\; \mathsf{True}\;|\; \mathsf{False}\;|\; (M,N)\; |\; \mathsf{fst }\; L \;|
\; \mathsf{snd } \;L\;|\;\lambda x.N \;|\; L\; M\\
&&& \mathsf{let} \; x = M \; \mathsf{in}\; N \; | \; \mathsf{if} \; L \; 
\mathsf{then} \; M \; \mathsf{else} \; N\\
\text{Environments}& \Gamma, \Delta & ::=& x_1:A_1, \ldots, x_n:A_n\\
\end{array}\\ 
\textbf{Types}\\
\begin{array}{ccc}
\prooftree
\justifies 
\emptyset \vdash \mathsf{False}:\mathbf{Bool}
\endprooftree&
\prooftree
\justifies
\emptyset \vdash \mathsf{True} : \mathbf{Bool}
\endprooftree &
\prooftree
(x:A)\in \Gamma
\justifies
\Gamma \vdash x :A
\endprooftree
\\ &&\\
\prooftree
\Gamma \vdash M : A \;\; \;\;\;\Gamma \vdash N : B
\justifies
\Gamma \vdash (M,N):A \times B
\endprooftree&
\;\;\;\;\;\;\prooftree
\Gamma \vdash L : A \times B
\justifies
\Gamma \vdash \mathsf{fst}\; L :A
\endprooftree
&
\;\;\;\;\;\prooftree
\Gamma \vdash L : A \times B
\justifies
\Gamma \vdash \mathsf{snd}\; L :B
\endprooftree
\end{array} \\ \\
\begin{array}{cc}
\;\;\;\;\;\;\;\;\prooftree
\Gamma, x : A \vdash N :B
\justifies
\Gamma \vdash \lambda x.N:A \to B
\endprooftree&
\;\;\;\;\;\;\;\;\;\;\;\prooftree
\Gamma \vdash L : A \to B \;\;\;\; \Gamma \vdash M :A
\justifies
\Gamma \vdash L\; M : B
\endprooftree \\ \\
\;\;\;\;\;\;\;\;\prooftree
\Gamma \vdash M : A \;\; \; \Gamma, x : A \vdash N :B
\justifies
\Gamma \vdash  \mathsf{let} \; x = M \; \mathsf{in}\; N  : B
\endprooftree&
\;\;\;\;\;\;\;\;\;\;\;\prooftree
\Gamma \vdash L : \mathbf{Bool} \;\;\;\; \Gamma \vdash M, N :B
\justifies
\Gamma \vdash \; \mathsf{if} \; L \; 
\mathsf{then} \; M \; \mathsf{else} \; N : B
\endprooftree
\end{array}
 \\\textbf{Laws}\\
\begin{array}{llcl}
(\beta_1^x) & \mathsf{fst}\;(M,N) &=& M\\
(\beta_2^x) & \mathsf{snd}\;(M,N) &=& N\\
(\eta^x)& (\mathsf{fst}\; L, \mathsf{snd} L) &=& L\\
(\beta^{\to})& (\lambda x.N)M &=& N[x :=M]\\
(\eta^{\to})& \lambda x. (L \;x) &=& L\\
(\mathsf{let})& \mathsf{let} \; x = M \; \mathsf{in}\; N &=& N[x := M]\\
(\beta_1^{\mathsf{if}}) &  \mathsf{if} \; \mathsf{True} \; \mathsf{then} \; M\; \mathsf{else} \; N &=& M\\
(\beta_2^{\mathsf{if}}) &  \mathsf{if} \; \mathsf{False} \; \mathsf{then} \; M\; \mathsf{else} \; N &=& N\\
\end{array}
\\\hline
\end{array}
\]
\end{figure}

\section{Monadic Calculus}
\label{app:monadic}

The simply-typed lambda calculus presented in Appendix~\ref{app:lambda} is
the foundation of purely functional programming languages.  In this section
we show the \emph{monadic calculus}~\cite{moggi89}, which also models monadic
effects. A monad is represented using a type constructor for computations
$m$ and two functions: $return :: a \to m \;a$ and $\bind :: m \;a \to (a \to
m \;b) \to m \;b$.  The operation $\bind$ (pronounced ``bind'') specifies how
to sequence computations and $return$ specifies how to lift values to
computations.  From a programming perspective, a \emph{monad} is a construct
to structure \emph{computations}, in a functional environment, in terms of
values and sequence of computations using those values.

The monadic calculus extends the simply-typed lambda calculus with the
constructs in Figure~7.  Unit and bind satisfy left unit,
right unit, and associativity laws, (left), (right), and (assoc).

\begin{figure}[!htb]
\label{fig:monadic}
\caption{Monadic Calculus}
\[
\begin{array}{l}\hline
\textbf{Syntax}\\ \begin{array}{llcl}
\text{Types} & A,B,C& ::=& ...\;|\;\mathbf{M} \; A \\
\text{Terms}& L,M,N & ::= & ...\; |\; [M]\;| \; \mathsf{let} \; 
x = M \; \mathsf{in} \; N \; |\; \mathsf{mzero} \; | \;\mathsf{+} \;|\; \mathsf{-}\\
\end{array}\\ 
\textbf{Monadic Types}\\
\begin{array}{cc}\\
\prooftree
\Gamma \vdash M : A
\justifies
\Gamma \vdash [M]: \mathbf{M}\; A
\endprooftree&
\;\;\;\;\;\;\prooftree
\Gamma \vdash M : \mathbf{M} \; A\;\;\;\; \Gamma, x : A \vdash N : \mathbf{M} \; B
\justifies
\Gamma \vdash \mathsf{let} \; x = M \; \mathsf{in} \; N : \mathbf{M}\; B
\endprooftree
\end{array} 
\\\textbf{MonadPlus Types}\\
\begin{array}{cc}\\
\prooftree
\justifies
\Gamma \vdash \mathsf{mzero} : \mathbf{M}\; A
\endprooftree&
\;\;\;\;\;\;\prooftree
\Gamma \vdash M,N : \mathbf{M} \; A
\justifies
\Gamma \vdash M + N : \mathbf{M}\; A
\endprooftree
\end{array} 
 \\\textbf{Laws}\\
\begin{array}{llcl}
(\text{left})&  \mathsf{let} \; x = [L] \; \mathsf{in} \; N&=& N[x:=L]\\
(\text{right})& \mathsf{let} \; x = L \; \mathsf{in} \; [x] &=& L\\
(\text{assoc})& \mathsf{let} \; y = (\mathsf{let} 
\; x = L \; \mathsf{in} \; N) \; \mathsf{in} \; T &=& \mathsf{let} 
\; x = L \; \mathsf{in} \; (\mathsf{let} 
\; y = N \; \mathsf{in} \; T)\\
\end{array}
\\\textbf{MonadPlus Laws}\\
\begin{array}{lcl}
\mathsf{mzero} + a &=& a\\
a + \mathsf{mzero} &=& a\\
a + (b + c) &=& (a + b) + c\\
\mathsf{let} \; x= \mathsf{mzero} \; \mathsf{in} \; T &=& \mathsf{mzero} \\
\mathsf{let} \; x =(M+N) \; \mathsf{in} \;T &=& 
(\mathsf{let} \; x = M \; \mathsf{in} \; T) + (\mathsf{let} \; x = N \; \mathsf{in} \; T )
\end{array}
\\\hline
\end{array}
\]
\end{figure}

Beyond the three monad laws discussed  above, some monads obey the \- \textsf{MonadPlus} laws. 
The \textsf{MonadPlus} interface provides two primitives, \textsf{mzero} and $+$ 
(called \textsf{mplus}), for expressing choices. The command $+$ introduces a
choice junction, and \textsf{mzero} denotes failure.


The precise set of laws that a \verb|MonadPlus| implementation should satisfy
is not agreed upon~\cite{HAWiki:MonadPlus}, but in~\cite{Hinze:deriving} is presented 
a reasonable agreement on the laws. We use in Figure~7 the laws introduced by~\cite{Hinze:deriving}.  


The intuition behind these laws is that \textsf{MonadPlus} is a disjunction of
goals and $\bind$ is a conjunction of goals. The conjunction evaluates the
goals from left-to-right and is not symmetric.

\section{General Quantum Computations}
\label{app:qcomp}

Quantum computation, as its classical counterpart, can be seen as processing
of information using quantum systems.  Its basic idea is to encode data using
quantum bits (qubits).  In quantum theory, considering a \emph{closed}
quantum system, the qubit is a \emph{unit} vector living in a complex inner
product vector space know as \emph{Hilbert space}~\cite{NielsenChuang00}.  We
call such a vector a \emph{ket} (from \emph{Dirac's notation}) and denote it
by $|v\rangle$ ( where $v$ stands for elements of an orthonormal basis), a
column vector.  Differently from the classical bit, the qubit can be in a
\emph{superposition} of the two basic states written as $\alpha |0\rangle+
\beta |1\rangle$, or

\[
\left(\begin{array}{c}
  \alpha\\
  \beta \\
\end{array}\right)
\]

\noindent 
with $|\alpha|^2 + |\beta|^2 = 1$.  Intuitively, one can think that a qubit
can exist as a $0$, a $1$, or simultaneously as both $0$ and $1$, with
numerical coefficient (i.e., the probability amplitudes $\alpha$ and $\beta$)
which determines the probability of each state.  The quantum superposition
phenomena is responsible for the so called ``quantum parallelism.''

Operations acting on those \emph{isolated} or \emph{pure} quantum states are
linear operations, more specifically \emph{unitary matrices} $S$. A matrix
$A$ is called \emph{unitary} if $S^* S=I$, where $S^*$ is the adjoint of $S$,
and $I$ is the identity.  Essentially, those unitary transformations act on
the quantum states by changing their probability amplitudes, without loss of
information (i.e., they are reversible).  The application of a unitary
transformation to a state vector is given by usual matrix multiplication.

Unfortunately in this model of quantum computing, it is difficult or
impossible to deal formally with another class of quantum effects, including
measurements, decoherence, or noise.

 
Measurements are critical to some quantum algorithms, as they are the only
way to extract \emph{classical} information from quantum states.

A \emph{measurement} operation projects a quantum state like $\alpha
|0\rangle+ \beta |1\rangle$ onto the basis $|0\rangle$,$|1\rangle$. The outcome of the
measurement is not deterministic and it is given by the probability
amplitude, i.e., the probability that the state after the measurement is $|0\rangle$
is $|\alpha|^2$ and the probability that  the state is $|1\rangle$ is
$|\beta|^2$.  If the value of the qubit is initially unknown, than there is
no way to determine $\alpha$ and $\beta$ with that single measurement, as the
measurement may \emph{disturb} the state.  But, \emph{after} the measurement,
the qubit is in a \emph{known} state; either $|0\rangle$ or $|1\rangle$.  In fact, the
situation is even more complicated: measuring part of a quantum state
collapses not only the measured part but any other part of the global state
with which it is \emph{entangled}.  In an entangled state, two or more qubits
have to be described with reference to each other, even though the
individuals may be spatially separated~\footnote{For more detailed
explanation about entangled, see~\cite{NielsenChuang00}.}.

There are several ways to deal with measurements in quantum computing, as
summarized in our previous work~\cite{viz06}. To deal formally and elegantly
with measurements, the state of the computation is represented using a
\emph{density matrix} and the operations are represented using
\emph{superoperators}~\cite{276708}. Using these notions, the
\emph{projections} necessary to express measurements become expressible
within the model.

Intuitively, density matrices can be understood as a statistical perspective
of the state vector.  In the density matrix formalism, a quantum state that
used to be modeled by a vector $|v\rangle$ is now modeled by its outer
product $|v\rangle \langle v|$, where $\langle v |$ is the row vector
representing the adjoint (or dual) of $|v\rangle$.  For instance, the state
of a quantum bit $|v\rangle = \frac{1}{\sqrt{2}}|0\rangle +
\frac{1}{\sqrt{2}}|1\rangle $ is represented by the density matrix:

\[
\left(\begin{array}{cc}
  \frac{1}{2} & -\frac{1}{2}\\
  -\frac{1}{2} & \frac{1}{2} \\
\end{array}\right)
\]
\noindent
Note that the main diagonal shows the classical probability distribution of
basic quantum states, that is, these state has $\frac{1}{2}$ of probability
to be $|0\rangle$ and $\frac{1}{2}$ of probability to be $|1\rangle$.

However, the appeal of density matrices is that they can represent states
other than the pure ones above. In particular if we perform a measurement on
the state represented above, we should get $|0\rangle$ with probability $1/2$
or $|1\rangle$ with probability $1/2$. This information, which cannot be
expressed using vectors, can be represented by the following density matrix:
\[
\left(\begin{array}{cc}
  1/2 & 0 \\
  0 & 0\\
\end{array}\right)
+ 
\left(\begin{array}{cc}
  0 & 0 \\
  0 & 1/2 \\
\end{array}\right) =
\left(\begin{array}{cc}
  1/2 & 0 \\
  0 & 1/2 \\
\end{array}\right)
\]

Such a density matrix represents a \emph{mixed state} which corresponds to
the sum (and then normalization) of the density matrices for the two results
of the observation.

The two kinds of quantum operations, namely unitary transformation and
measurement, can both be expressed with respect to density
matrices~\cite{selinger06}.  Those operations now mapping density matrices to
density matrices are called \emph{superoperators}. A unitary transformation
$S$ maps a pure quantum state $|u\rangle$ to $S|u \rangle$. Thus, it maps a
pure density matrix $|u\rangle \langle u|$ to $S |u\rangle \langle u| S^*$.
Moreover, a unitary transformation extends linearly to mixed states, and
thus, it takes any mixed density matrix $A$ to $S A S^*$.

As one can observe in the resulting matrix above, to execute a measurement
corresponds to setting a certain region of the input density matrix to zero.

\section{Definition of Measurement Operations}
\label{app:measop}

In this section we present the denotations of the programs for 
measurements, $trl$ and $meas$, added to the quantum arrow calculus.

\[
\begin{array}{l}
\mathsf{trL} :: \mathbf{Super} \;(A,B) \;B \\
\mathsf{trL} ((a_1,b_1),(a_2,b_2)) =  \mathsf{if} \;a_1 == a_2 
\; \mathsf{then} \; return (b_1,b_2) \; \mathsf{else} \;\mathsf{mzero} \\
\\
\mathsf{meas} :: \mathbf{Super} \;A \;(A,A) \\
\mathsf{meas} (a_1,a_2) = \mathsf{if}\; a_1 == a_2 \; \mathsf{then} \; return ((a_1,a_1),(a_1,a_1)) \; 
\mathsf{else} \; \mathsf{mzero}
\end{array}
\]

\noindent
We consider \emph{projective}  measurements which are 
described by a set of projections onto mutually 
orthogonal subspaces. This kind of measurement returns a classical  
value and a post-measurement state of the quantum system.
The operation $meas$ is defined in such a way that it can
encompass both results. Using the fact that a classical 
value $m$ can be represented by the density 
matrix $|m \rangle \langle m|$ the superoperator $meas$ 
returns the output of the measurement attached to the 
post-measurement state. 

\end{document}